\documentclass[%
 aip,
 jap,%
 amsmath,amssymb,
preprint,%
]{revtex4-1}

\usepackage{graphicx}
\usepackage{dcolumn}
\usepackage{bm}
\usepackage{epsfig}

\begin{document}

\title{A Continuous Cold Atomic Beam Interferometer}
\author{Hongbo Xue}
\affiliation{State Key Laboratory of Precision Measurement Technology and Instruments, Tsinghua University, Beijing 100084, P. R. China}
\affiliation{Joint Institute for Measurement Science, Tsinghua University, Beijing 100084, P. R. China}
\affiliation{State Key Laboratory of Space Weather, National Space Science Center, Chinese Academy of Sciences, Beijing 100190, P. R. China}

\author{Yanying Feng}
\thanks{Author to whom correspondence should be addressed}
\email{yyfeng@tsinghua.edu.cn}
\affiliation{State Key Laboratory of Precision Measurement Technology and Instruments, Tsinghua University, Beijing 100084, P. R. China}
\affiliation{Joint Institute for Measurement Science, Tsinghua University, Beijing 100084, P. R. China}

\author{Shu Chen}
\affiliation{Joint Institute for Measurement Science, Tsinghua University, Beijing 100084, P. R. China}
\affiliation{Key Laboratory of Instrumentation Science, North University of China, Taiyuan 030051, P. R. China}

\author{Xiaojia Wang}
\affiliation{College of Mechanical Engineering, Taiyuan University of Technology, Taiyuan 030024, P. R. China}

\author{Xueshu Yan}
\affiliation{State Key Laboratory of Precision Measurement Technology and Instruments, Tsinghua University, Beijing 100084, P. R. China}
\affiliation{Joint Institute for Measurement Science, Tsinghua University, Beijing 100084, P. R. China}

\author{Zhikun Jiang}
\affiliation{State Key Laboratory of Precision Measurement Technology and Instruments, Tsinghua University, Beijing 100084, P. R. China}
\affiliation{Joint Institute for Measurement Science, Tsinghua University, Beijing 100084, P. R. China}

\author{Zhaoying Zhou}
\affiliation{State Key Laboratory of Precision Measurement Technology and Instruments, Tsinghua University, Beijing 100084, P. R. China}

\begin{abstract}
We demonstrate an atom interferometer that uses a laser-cooled continuous beam of $^{87}$Rb atoms having velocities of 10--20~m/s. With spatially separated Raman beams to coherently manipulate the atomic wave packets, Mach--Zehnder interference fringes are observed at an interference distance of 2L = 19~mm. The apparatus operates within a small enclosed area of 0.07~mm$^2$ at a bandwidth of 190~Hz with a deduced sensitivity of $7.8\times10^{-5}$~rad/s/$\sqrt{\mbox{Hz}}$ for rotations. Using a low-velocity continuous atomic source in an atom interferometer enables high sampling rates and bandwidths without sacrificing sensitivity and compactness, which are important for applications in real dynamic environments.

\begin{description}
\item[PACS numbers]
37.25.+k, 32.80.Qk, 37.10.Jk, 05.60.Gg
\end{description}

\end{abstract}\maketitle

\section{Introduction}
Since their inception in 1991 \cite{riehle_optical_1991,kasevich_atomic_1991}, light-pulse atom interferometers (LPAIs) have shown the potential to be extremely sensitive sensors in many applications such as gravimeter surveys \cite{peters_high-precision_2001,yu_development_2006, bodart_cold_2010, hu_demonstration_2013}, seismic studies \cite{mcguirk_sensitive_2002}, inertial navigation \cite{canuel_six-axis_2006,durfee_long-term_2006,wu_demonstration_2007,dickerson_multiaxis_2013}, tests of fundamental physics \cite{muller_precision_2010}, and measurements of fundamental constants \cite{cadoret_combination_2008}. From the early developmental stages of atom interferometer, different thermal atomic beam interferometers have been constructed with mechanical gratings \cite{keith_interferometer_1991,lenef_rotation_1997} or light gratings \cite{rasel_atom_1995, tonyushkin_magnetic_2010} for the coherent manipulation of atomic waves. Among them, a LPAI-based gyroscope using a thermal atomic beam has exhibited excellent performances in short-term noise, long-term stability, and bandwidth ($\sim$ 110 $Hz$) \cite{gustavson_rotation_2000,durfee_long-term_2006}. For a thermal atomic beam interferometer gyroscope with a high longitudinal atomic velocity (220–-300~m/s), it is a challenge to reduce the system’s dimension without sacrificing sensitivity.
 
Using cold atoms is a natural solution towards constructing a compact atomic interferometer, in which a pulse-launched cold atom cloud is usually used as a matter-wave source \cite{wang_demonstration_2007,gauguet_characterization_2009,tackmann_self-alignment_2012,dickerson_multiaxis_2013}. In more dynamic environments, the pulsed-source-based LPAIs in their current forms may not complement or replace conventional technologies because of their relatively long cycle time and therefore low data rate, being the reciprocal of the interferometer cycle time. To date, most reported cold LPAIs demonstrations have operated at a data rate of a few Hertz or less. Although a combination of conventional and atom interferometer technologies may be a promising solution to actual applications of such cold LPAIs \cite{merlet_operating_2009,geiger_detecting_2011,aaron_canciani_integration_2012}, significant advances in LPAI performance or data-rate improvements of order 100~Hz or more are required.

Another method is the short interrogation LPAI or high data rate LPAI \cite{butts_light_2011,mcguinness_high_2012}, which trade sensitivity for data rate and reduced system demands. However, rather than trading off data rate and sensitivity using short interrogation LPAIs with a pulsed cold atomic source, our method consists in eliminating the dead times (time intervals without atoms in the interference region) by using a continuous beam of cold atoms. A continuous atomic source also helps to reduce noise induced by inter-modulation effects and collision shifts \cite{guena_experimental_2007,devenoges_improvement_2012}.
An atomic fountain clock with a continuous cold atomic beam\cite{di_domenico_uncertainty_2011} has been demonstrated with a stability of $6\times 10^{-14}$ $/\sqrt{\tau}$, which gives a good solution to challenges provided by a continuous cold atomic source, such as shielding the fluorescence from the atomic source, compared with its pulsed counterpart.
In this paper, we describe an atom interferometer, which uses a continuous cold atomic beam as source and spatial-separated Raman pulses for the coherent manipulation of atomic wave-packets. A continuous atomic beam enables higher bandwidths from an atom interferometer when seeking similar sensitivity and compactness as an atom interferometer using a pulsed cold atom source.

\section{Setup}

\begin{figure*}
\graphicspath{}
\centerline{\psfig{file=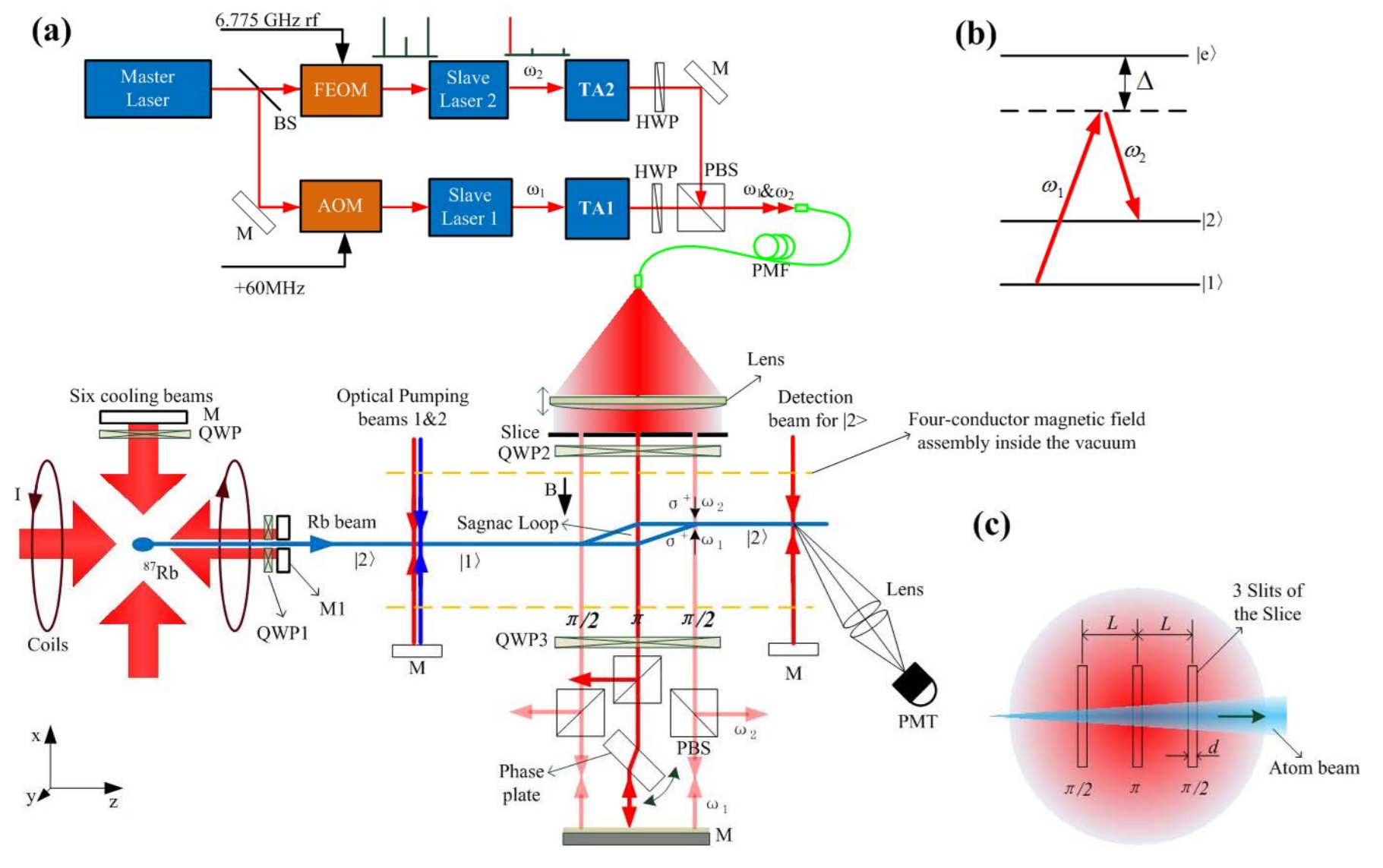,width=1\linewidth}}
\caption{(a) Schematic of the experimental layout for the atom interferometer based on a laser-cooled $^{87}Rb$ atomic beam. BS: beam splitter; M: mirror; HWP: half-wave plate; QWP: quarter-wave plate; PMF: polarizing maintained fiber; FEOM: fiber electro-optic modulator; TA: tapered diode laser amplifier; PMT: photomultipliers. (b) Level diagram for the ground states $|1\rangle$ and $|2\rangle$ and the excited state $|e\rangle$. The frequencies $\omega_1$ and $\omega_2$ are used to induce Raman transitions between the two ground states. (c) The $\pi/2$--$\pi$--$\pi/2$ pulse sequence is realized by blocking a Gaussian beam with slit width of $d = 1.0$~mm and interval of $L = 9.5$~mm.}
\label{fig:1}
\end{figure*}

A diagram of the experimental setup is shown in Fig.~\ref{fig:1}. The vacuum system consists of two chambers, each of which is pumped with a single ion pump. One chamber, working for an imbalanced three-dimensional (3D) magneto-optical trap (MOT), is pumped to $2 \times 10^{-9}$ Torr, and the other chamber, working for the interferometer, is pumped to $8 \times 10^{-10}$ Torr when the Rubidium reservoir is turned off. The two chambers are separated by a $\lambda/4$ plate and a mirror, denoted as QWP1 and M1, respectively, each of which has a hole of diameter 1~mm at its center (Fig.~\ref{fig:1}). A low-velocity intense source (LVIS) of cold $^{87}$Rb atoms is generated from one chamber in which cold atoms are prepared in a vapor-cell 3D MOT\cite{lu_low-velocity_1996, wang_cold_2011}. Standard 3D MOT optics with the retro-reflective configuration are implemented with the cooling light tuned to the $^{87}$Rb $5s^2S_{1/2}$, F=2 $\rightarrow$ $5p^2P_{3/2}$, $F^{\prime}=3$ allowed transition with a typical red detuning of $\delta=$4--5$\Gamma$ (where $\Gamma=2\pi \times 6$~MHz is the natural linewidth of $^{87}Rb$) and the repumping light tuned to the $5s^2S_{1/2}, F=1 \rightarrow 5p^2P_{3/2}, F^{\prime}=2$ allowed transition. To trap the atoms, a pair of anti-Helmholtz coils oriented along the z-axis is used to generate a quadrupole magnetic field with a transverse field gradient of about 15 G/cm. Six 45~mm waist beams of the 3D MOT have a 4.5~mW/cm$^2$ peak intensity in the $\sigma ^{+} / \sigma^{-}$ configuration. Using the holes drilled in QWP1 and M1, one pair of cooling beams is produced along the z-axis by retro-reflection of the light beam. This generates a dark channel along the z-axis and allows cold atoms to leak out continuously from the MOT chamber because of the unbalanced radiation pressure. The geometric axis of the MOT chamber has an angle of 3.7 $^{\circ}$ with respect to that of the interferometer chamber, which is placed horizontally. Hence, the cold atomic beam is launched at the same angle with respect to the horizontal direction, and separated from the leaking cooling light beam through the effect of gravity. The leaking cooling light and fluorescence from the atomic source are absorbed by the inner walls of the interferometer chamber, which are coated with a layer of black lacquer.

The most-probable longitudinal velocity of the atomic beam can be tuned from 10~m/s to 20~m/s and the atomic flux set above $4\times10^9$~atoms/s. After their continuous extraction from the 3D MOT, the cold atoms are pumped into a single hyperfine $F=1$ level using optical pumping laser 1 (Fig.~\ref{fig:1}), which is tuned to the $F=2 \rightarrow F^{\prime}=2$ allowed transition. The optical pumping laser 2, tuned to $F=1 \rightarrow F^{\prime}=0$ in $\sigma^{\pm}$ transitions (linear polarization orthogonal to the magnetic field), is used to drive atoms from the F = 1, $m_F =\pm 1$ levels to the F = 1, $m_F = 0$ magnetic field insensitive ground-state hyperfine level. The continuous cold atomic beam then enters the interferometer interaction region with no further cooling or collimation.

Two-photon velocity-selective Raman transitions are used to manipulate the atomic wave packets, while keeping them in long-lived ground states. The optical apparatus of the Raman lasers is shown in Fig.~\ref{fig:1}. The master laser is a 780~nm external cavity diode laser (ECDL, typical linewidth of 300~kHz, DL Pro, Toptica, Germany), which is frequency-stabilized using a high finesse wavelength meter (WS6-200, Toptica) tuned to the $F=1 \rightarrow F^{\prime}=1$ allowed transition with a large red detuning of $\Delta = – 2\pi \times 1.07$~GHz. The two slave lasers are GaAsP/AlGaAs Fabry-Perot CW diode lasers. Slave laser 1, operating at frequency $\omega_1$, is injection-locked to the frequency-shifted master laser by an acousto-optic modulator (AOM) driven by a 60~MHz signal. Another master laser beam passes through a fiber electro-optical modulator (FEOM; EOspace, USA), driven by a RF signal with frequency $\omega_{00}- 2\pi \times$ 60~MHz = $2\pi \times 6.775$~GHz, where $\omega_{00}=\omega_1-\omega_2$ = $2\pi \times$6.835~GHz corresponds to the ground-state hyperfine transition of $^{87}Rb$. Slave laser 2 is synchronized with the -1 sideband of the FEOM output by the frequency-selective sideband injection-locking technique \cite{xue_note:_2013}. Both the 6.775~GHz and 60~MHz RF sources are referenced to the same Cs atomic clock. The beat frequency between $\omega_1$ and $\omega_2$ was observed by mixing outputs of the two slave lasers on a fast photodiode (Fig.1(a)). The -3~dB line-width of the beat frequency is about 1.5~Hz when the 60~MHz AOM was turned on, but is less than 1~Hz, which is the resolution of the spectrum analyzer, when the AOM was turned off. The line-width worsens because of noise induced by the AOM driving signal. The single side band phase noise spectrum was -54~dBc/Hz at 3~Hz offset and about -105~dBc/Hz in the 3-32~kHz offset range. Beams from slave lasers 1 and 2 are power-amplified to approximately 600~mW, each using a semiconductor laser amplifier (TA1 and TA2, BoosTA, Toptica). These two beams are then combined into a single-mode polarization-maintaining fiber with crossed polarizations and expanded to 60~mm waist beams by a Gaussian doublet fiber coupler. Then, the Gaussian beam is blocked by a slice with three parallel slits with a spacing of $L = 9.5$~mm and a width of $d = 1.0$~mm, which allows power intensities of the two $\pi /2$ Raman pulses to be half that of the $\pi$ Raman pulse in the center of the Gaussian beam (Fig.~\ref{fig:1}(c)). By adjusting the distance between the lens (f = 220~mm) and the fiber, the two $\pi / 2$ Raman pulse beams, with equal optical paths, can be made parallel with the $\pi$ pulse. Three Raman pulse beams with crossed linear polarization are converted into opposite circular polarizations using a quarter-wave plate (QWP2). After passing through the atomic beam, the light fields are linearized by another quarter-wave plate (QWP3), and $\omega_1$ and $\omega_2$ are spatially separated by a polarizing beam splitter (PBS). Only the $\omega_1$ beam is retro-reflected back through the atomic beam. The counter-propagating Raman laser beams are generated for Doppler-sensitive Raman transitions in the form of $\sigma ^+$--$\sigma ^+$ or $\sigma^-$--$\sigma ^-$ transitions. The three Raman beams are set on the same optical table and could be rotated relative to the atomic beam. Removing the QWP2 and blocking the retro-reflected beam allows Raman transitions in Doppler-insensitive geometries.

Interference fringes from the cold atomic beam are observed by varying the optical phase of the $\pi$ Raman pulse. As indicated in Fig.~\ref{fig:1}(a), this is achieved by tilting a 9.53-mm-thick optically flat phase plate, which is located initially at an angle of 45 $^{\circ}$ with respect to the $\pi$ Raman beams and rotated by a piezo-electric transducer (PZT, Pst 150/4/7 bs, Piezomechanik GmbH). Tilting the phase plate leads to a phase shift of only the $\omega_1$ beam as only this beam is retro-reflected back to the atomic beam. The relative phase shift of the $\pi$ Raman beams and therefore the atomic phase shift can vary by changing the optical path length of $\omega_1$ through the phase plate when scanning the PZT. To detect the atomic signals, e.g. the interference fringes, the fluorescence of the F = 2 state atoms induced by the detection laser beam ($F=2 \rightarrow F^{\prime}=3$) is continuously collected by a photomultiplier tube (PMT, H7422-50, Hamamatsu, Japan). A horizontal magnetic bias field in the direction of the Raman beams is generated throughout the length of the interferometer interaction region, with a four-conductor magnetic field assembly running inside the main vacuum chamber \cite{morris_shielded_1984}, as shown in Fig.~\ref{fig:1}. Currents of about 4~A are applied to generate the bias field strength of 0.96~Gauss, which corresponds to a Zeeman frequency shift of 0.67~MHz and is sufficient to separate the Zeeman sublevel energies apart by an amount comparable to the Doppler-sensitive Raman transition linewidth or about 500~kHz, and allows the Raman transitions address only the $m_F = 0$ atoms.
\section{Results}

\begin{figure}
\graphicspath{}
\centerline{\psfig{file=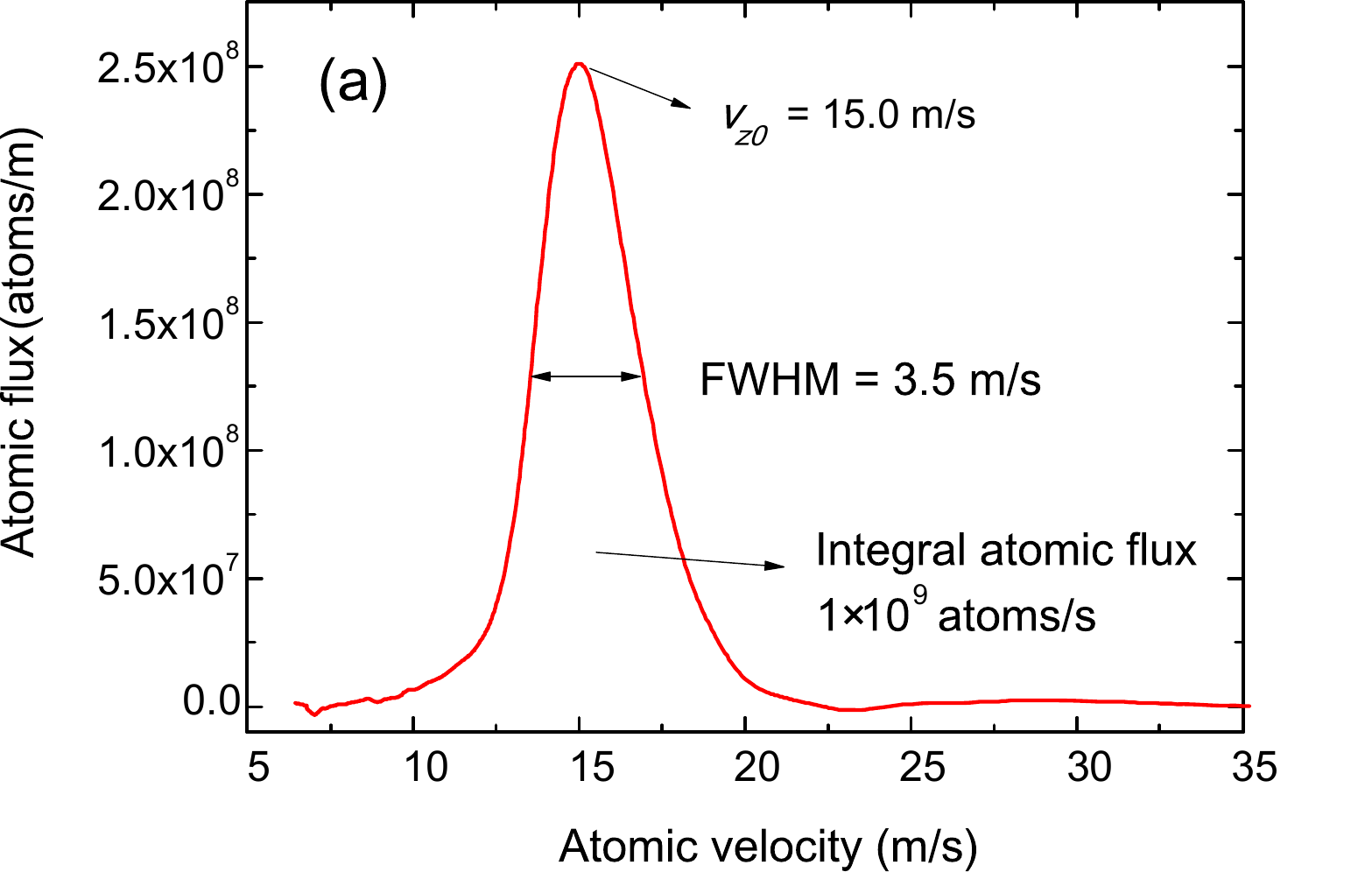,width=0.7\linewidth}}
\centerline{\psfig{file=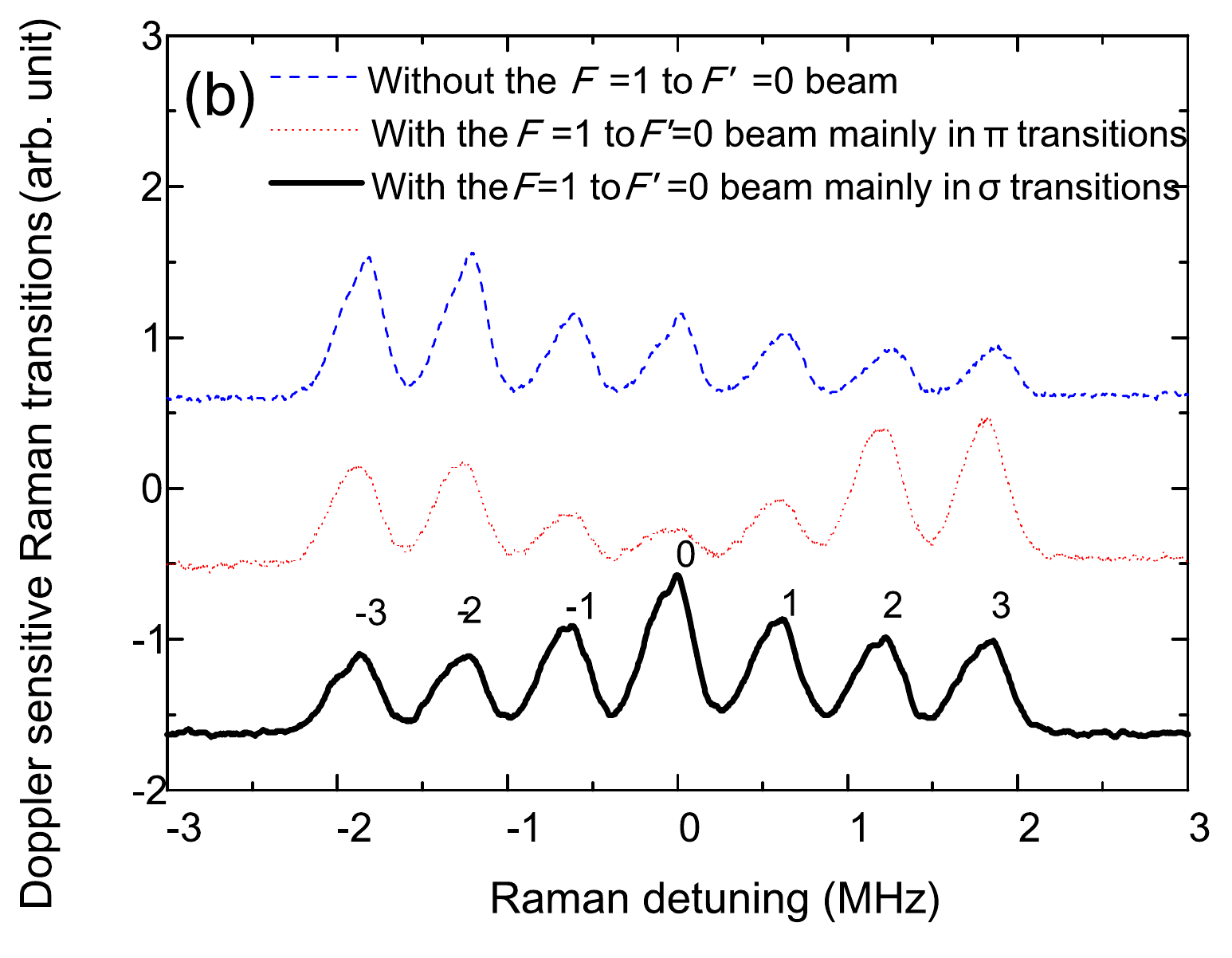,width=0.7\linewidth}}
\caption{Measurement of the cold atomic beam. (a) The longitudinal velocity distribution of the cold atomic beam measured using the time of flight method (TOF). (b) The state preparation of $|F=1, m_F=0\rangle$, which is indicated by the Raman transition spectra driven by a $\pi$ Raman pulse. Peak 0: $m_F$=0 to $m_F$=0 Raman transition; Peaks $\pm 1$: $m_F$=0 to $m_F$=$\pm 1$ and $m_F$=$\pm 1$ to $m_F$=0 Raman transitions; Peaks $\pm 2$: $m_F$=+1 to $m_F$ =+1 and $m_F$=$-1$ to $m_F$=$-1$ Raman transitions, respectively; Peaks $\pm 3$: $m_F$=+1 to $m_F$ =+2 and $m_F$=$-1$ to $m_F$=$-2$ Raman transitions.}
\label{fig:2}
\end{figure}

The velocity and flux of the MOT-based cold atomic beam are measured by the time-of-flight (TOF) method, which is realized by plugging the atomic beam and simultaneously recording the falling edge of the fluorescence signal of atoms in the $F=2$ state. The spectrum of the atomic velocity is deduced from the TOF signal (Fig.~\ref{fig:2}(a)). The most-probable longitudinal velocity of the atomic beam is adjusted to $v_{z0}=15.0$~m/s, with a velocity distribution of $\delta v_z=\sim$ 3.5~m/s (FWHM) by changing the polarizations of the cooling lasers and the magnetic field gradient.

After launch from the 3D MOT, the atoms are optically pumped to the magnetically insensitive F=1, $m_F$=0 sub-level by applying the magnetic bias field and the two retro-reflected beams, one resonant with the $F=2 \rightarrow F^{\prime}=2$ transition with circular polarization and the other with the $F=1 \rightarrow F^{\prime}=0$ transition with linear polarization parallel to the bias field. The velocity-sensitive Raman transition spectra can be used for measuring the population of atoms in different ground-state hyperfine $m_F$ sub-levels. In our experiment, seven Raman peaks instead of three peaks occur because the net bias field is not well aligned with the Raman beams, being perturbed by Earth’s and environmental magnetic fields, and hence allows both $\pi$ and $\sigma$ Raman transitions to occur. This influences the quantitative measurement of the state preparation efficiency. In Fig.~\ref{fig:2}(b), the Raman peaks labeled 0 correspond to the magnetically insensitive $m_F$=0 to $m_F$=0 Raman transition; those labeled $\pm 1$ correspond to the magnetically sensitive $m_F$=0 to $m_F$=$\pm 1$ and $m_F$=$\pm 1$ to $m_F$=0 Raman transitions, respectively. Similarly, $\pm 2$ correspond to the $m_F$=+1 to $m_F$ =+1 and $m_F$=$-1$ to $m_F$=$-1$ transitions, whereas $\pm 3$ correspond to the $m_F$=+1 to $m_F$ =+2 and $m_F$=$-1$ to $m_F$=$-2$ transitions. For the blue-dash line of the same figure, atoms are asymmetrically distributed in all ground-state hyperfine $m_F$ levels with only the $F=2 \rightarrow F^{\prime}=2$ optically pumping beam. With the other optically pumping beam resonant with the $F=1 \rightarrow F^{\prime}=0$ transition, more atoms are prepared in the magnetically insensitive $F = 1$, $m_F = 0$ sub-level from the other magnetically sensitive $m_F$ sub-levels, as seen from the black-solid line in Fig.~\ref{fig:2}(b). Orientation of the linear polarization of the $F=1 \rightarrow F^{\prime}=0$ optically pumping beam needs to be fine-tuned to allow only the $\sigma$ transition to occur, otherwise the state-preparation efficiency will be reduced significantly with the presence of the $\pi$ transition, as seen from the red-dotted line in Fig.~\ref{fig:2}(b). The state-preparation efficiency can be evaluated roughly from the relative Raman peak amplitudes.

\begin{figure}
\centerline{\psfig{file=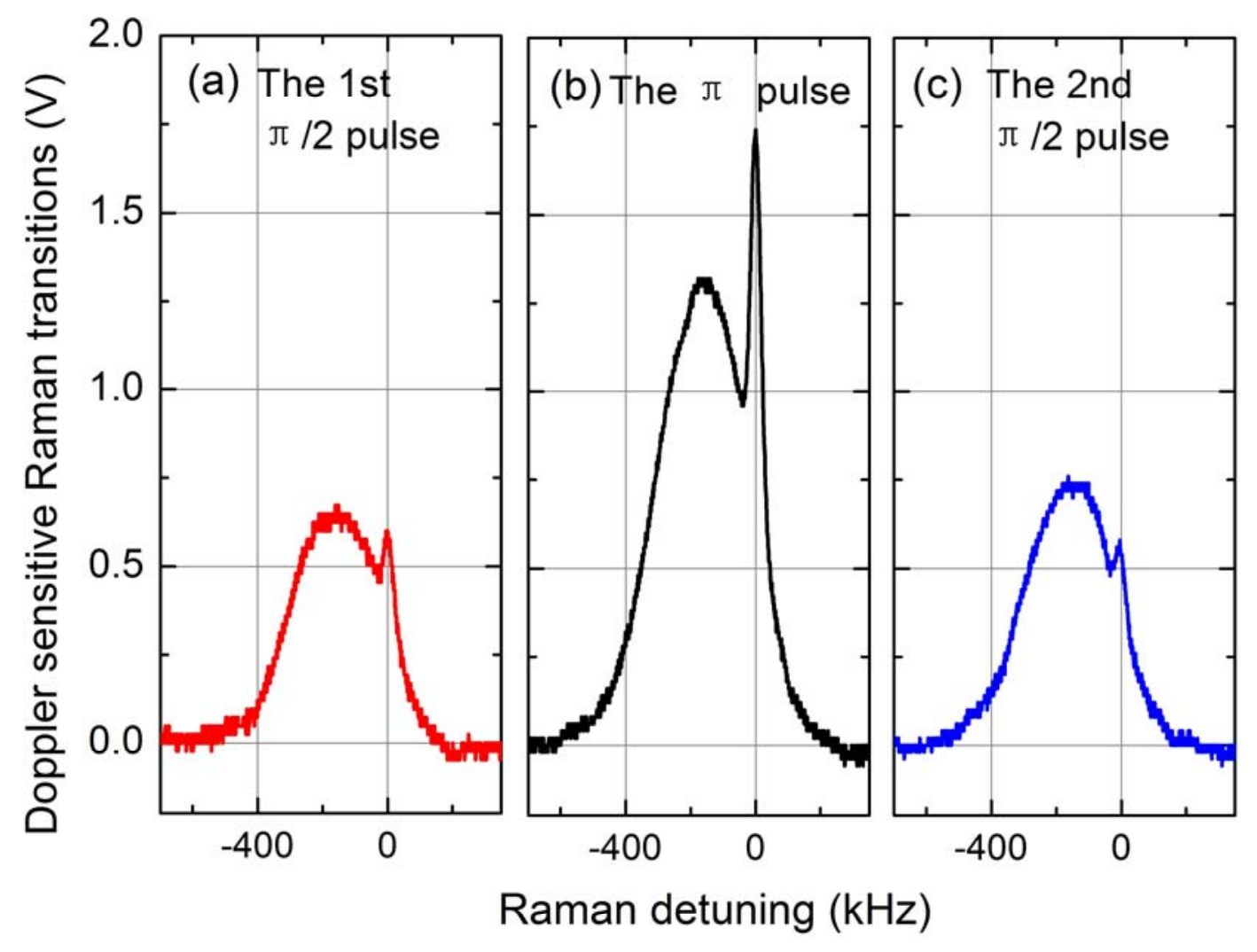,width=0.8\linewidth}}

\caption{Doppler-sensitive Raman transition spectra driven by the $\pi/2$--$\pi$--$\pi/2$ Raman pulse sequence. Here the longitudinal velocity of the atomic beam is $v_{z0}=15.0$~m/s; the longitudinal velocity spread is $\delta v_z=$ 3.5~m/s; the width of the Raman beams are d = 1.0~mm; the width of the transverse velocity distribution is $\Delta v_{x,beam}=\pm$ 6.5~cm/s. Doppler-insensitive Raman transitions were driven by the residual co-propagating Raman lasers, and were shifted from the peaks by tilting the Raman beams from the orthogonal propagation.}
\label{fig:3}
\end{figure}

The continuous cold atomic beam is manipulated by spatially separated Raman beams with constant pulse widths of d = 1.0~mm. The duration time $\tau = d/v_{z0}$ remains constant for each interaction of the atoms with the Raman beams. The Rabi frequency $\Omega_{eg}$ is adjusted by varying the intensities of the Raman beams of $\omega_1$ and $\omega_2$, and therefore the Rabi phase $\phi = \Omega_{eg} \tau$ can be set to $\pi /2$ or $\pi$. The corresponding spectra driven by the Doppler-sensitive $\pi/2$--$\pi$--$\pi/2$ Raman pulse sequence are shown in Fig.~\ref{fig:3}. The peak amplitudes of the spectra driven by the two $\pi/2$ pulses are equal and are half of that driven by the $\pi$ pulse. By tilting the Raman beams slightly away from the perpendicular direction with respect to the atomic beam, the Doppler-sensitive transition peaks are shifted away from peaks of the residual Doppler-insensitive transitions due to the impure polarization of the Raman beams. Doppler shifts and recoil shifts are compensated by offsetting the difference $\omega_1 -\omega_2$. The transverse velocity spread of the atomic beam was approximately $\pm$ 6.5~cm/s (FWHM), which can be estimated from the $\sim$ 336~kHz linewidth of the velocity-sensitive Raman transition spectrum driven by the $\pi$ Raman pulse ($\tau =67$~$\mu s$).

In implementing the continuous cold atom beam interferometer, stringent requirements are placed on the parallel alignment of the spatially separated Raman beams. This is because slow thermal beam diverge more than fast thermal beams under the same transverse equivalent temperature \cite{gustavson_rotation_2000}. The Raman beams should be aligned precisely before an interference signal can be seen. With bad beam alignment, the three spatially separated Raman beams will interact with atoms of different transverse velocities. For the horizontal alignment, the Doppler shift in the Raman transition, $-k_{eff}v_zsin \theta$, must be smaller than the linewidth of the Raman resonance, $\Delta \delta_{12} \approx 2\pi \frac{0.8}{\tau} = 2\pi\frac{0.8v_{z0}}{d}$, where $k_{eff}$ is the effective wave vector. This implies $\theta \ll$ 312~$\mu rad$, when the atomic longitudinal velocity is $v_{z0}$ = 15~m/s and the pulse width of Raman beams is d = 1.0~mm. By adjusting the distance between the lens and the fiber, and monitoring the feedback light into the fiber, we can achieve a horizontal beam alignment of $\theta < $ 91~$\mu rad$ between the three Raman beams (The fiber core is 5~$\mu m$ in diameter and NA = 0.11). The vertical Raman beam alignment is equally critical. It can be assumed that the three beams generated from the same fiber coupler are substantially parallel in the vertical direction. 

\begin{figure}
\graphicspath{}
\centerline{\psfig{file=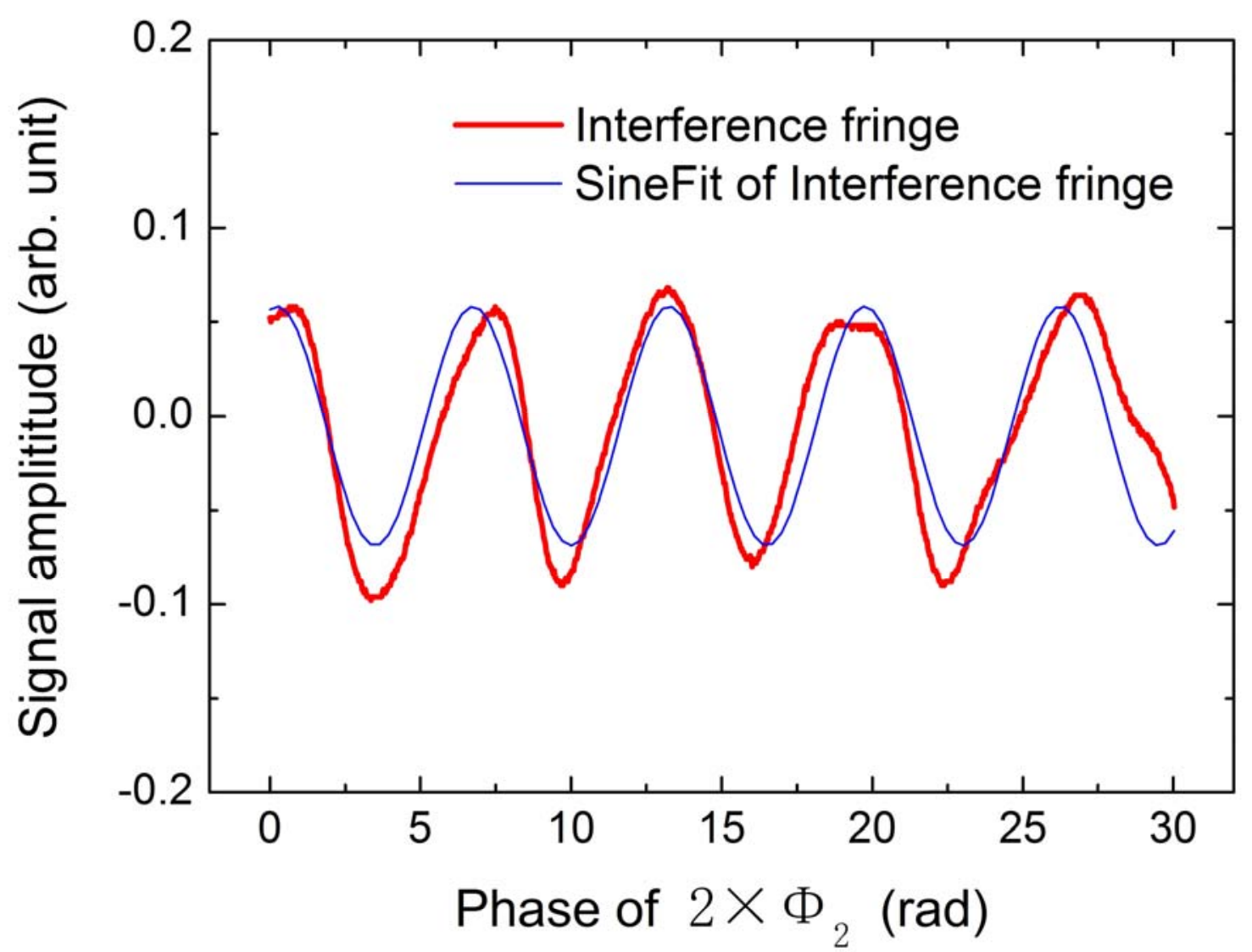,width=0.8\linewidth}}

\caption{The $\pi/2$--$\pi$--$\pi/2$ Mach--Zehnder atom interference fringe (red solid line) as a function of the optical phase of the $\pi$ Raman pulse, obtained by rotating the phase-plate in the optical path of one of the Raman beams.}

\label{fig:4}
\end{figure}

The signal of the $\pi/2$--$\pi$--$\pi/2$ Mach--Zehnder atom interferometer is written
\begin{equation}
P_e=\frac{1}{2} [1+cos(\Phi _a +\Phi _{\Omega} +\phi _1 -2\phi _2 +\phi _3)]
\end{equation}
where $\Phi_a$ is the phase shift caused by acceleration; $\Phi _{\Omega}$ is the Sagnac phase shift caused by rotation; $\phi _1$, $\phi _2$ and $\phi _3$ are effective phases of the first $\pi /2$, $\pi$, and the second $\pi /2$ Raman pulses, respectively. Interference fringes are observed by scanning the phase of the $\pi$ Raman pulse with the phase plate. The interference fringe as a function of $2\phi _2$ (Fig.~\ref{fig:4}) explicitly verifies that the phase shift of the interferometer output is twice that of the $\pi$ Raman pulse. The maximum extension of the PZT used for scanning the phase plate is 9~$\mu m$, which corresponds to a $\sim$ 30~rad phase shift of the interference signal. This can be determined by the $\pi$ Raman pulse phase shift as a function of the rotation angle of the phase plate. This phase modulation technique may be used for compensating and modulating the phase of the interferometer signal.

\begin{figure}
\centerline{\psfig{file=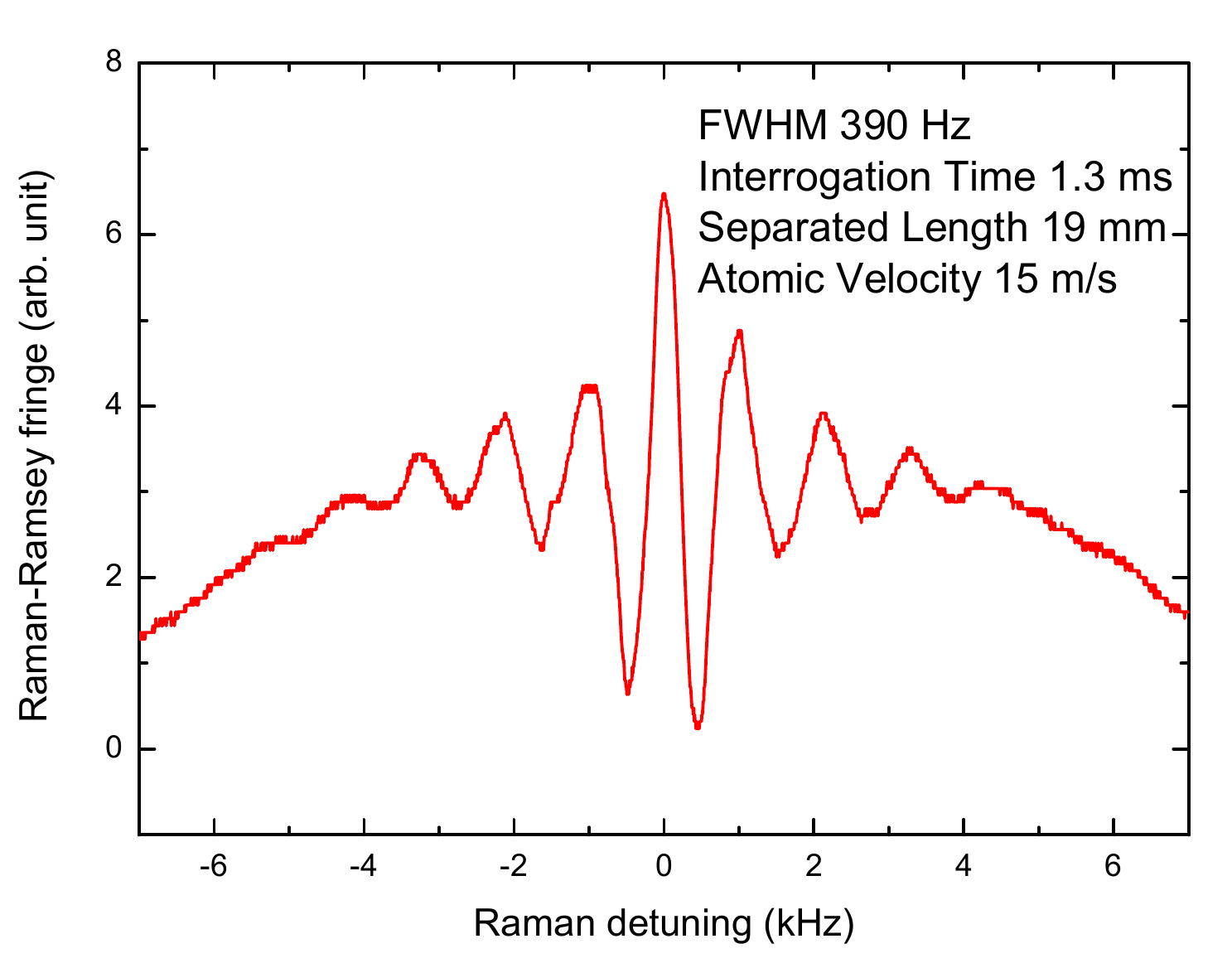,width=0.8\linewidth}}
\caption{Raman--Ramsey fringes driven by $\pi/2$--$\pi/2$ Raman pulses. The length of the separated oscillatory fields is L=19~mm and the width of the interaction zone is d=1.0~mm.}
\label{fig:5}
\end{figure}

There are several causes limiting the interference fringes contrast. First, the longitudinal velocity spread of the atomic beam is primarily responsible for the reduction in contrast. Phases of the Raman pulses ($\pi$ and $\pi /2$) should be set optimally for the most-probable longitudinal velocity $v_{z0}$ of the atomic beam. The Raman--Ramsey fringes from the cold atomic beam were observed (Fig.~\ref{fig:5}). These were obtained using the two $\pi/2$ Raman pulses in the Doppler-insensitive geometry (co-propagating beams), with a field separation length of 2L = 19~mm when the $\pi$ pulse was blocked. The atomic interrogation time between the two $\pi/2$ pulses was T = 1.3~ms, and the corresponding linewidth of the central fringe was $\delta_{Ramsey}$ = 1/(2T) = 390~Hz. Orders higher than 4$^{th}$-order of the interference fringe, where $n = v_{z0} / \delta v_z= \sim 4 $, can be identified clearly against the velocity average \cite{dickerson_multiaxis_2013}. This means that a cold atomic beam with a narrow longitudinal velocity distribution may increase the number of observable fringes of an atom interferometer, compared with its thermal counterpart ($n=\sim 1$)\cite{gustavson_rotation_2000}.

Second, the transverse velocity spread of the atomic beam will further decrease the contrast. When using Doppler-sensitive Raman pulses (counter-propagate Raman beams), atoms in the range of $\Delta v_x =\delta_{12}/k_{eff}$ are selected to interact with Raman pulses. In our experiments, the linewidth of the Raman transition is 12~kHz and the selected velocity range is about $\Delta v_x$ = 5~mm/s, when the pulse width of the Raman beams is d = 1.0~mm. This selected velocity range for joining the interference is much less than the measured transverse velocity spread ($\pm$ 6.5~cm/s) of the cold atomic beam. Atoms beyond this selected velocity range cannot be manipulated effectively by the Raman pulses and both contrast and signal-to-noise ratio of the interference signal deteriorate. In the present experiment, the peak to peak signal amplitude in Fig.~\ref{fig:4} corresponds to about $10^{7}$~atoms/s with 17$\%$ contrast. To increase the velocity-selected range of a cold atomic beam interferometer is a big challenge. One method to achieve a larger range is to increase the linewidth of the Raman transition by decreasing the pulse width d of the Raman beams.
 
Furthermore, because of the transverse velocities, the cold atomic beam will diverge after the atoms have traveled for an extended period. This thermal expansion of an atomic beam is extremely undesirable when manipulating the atomic beam with the Raman beams. This gives rise to inhomogeneous Rabi frequencies when atoms in different transverse positions interact with Raman beams that have spatially inhomogeneous intensity distributions, or when similar interactions occur between the three individual Raman beams. Consequently, the coherence between the Raman pulses and the atomic beam will be destroyed by this inhomogeneous average effect. Further sub-Doppler transverse cooling of the atomic beam can be performed to suppress the transverse thermal expansion of the cold atomic beam.

Before finally optimizing the horizontal alignment of the counter-propagating Raman beams and the Raman detuning, the relative ac Stark shift needs to be cancelled by carefully setting the intensity ratio, $I_2 /I_1$, between the two Raman beams ($\omega_1$ and $\omega_2$), otherwise an additional phase offset is introduced into the interferometer, which leads to a reduction in contrast and even vanishing of the interference signal when the nonzero relative ac Stark shift is larger than the linewidth of the Raman transition. Experimentally, we first tune the intensity ratio, $I_2 /I_1$, to have the Raman--Ramsey fringe centered on the Rabi flop of a single Raman pulse and then optimize the parameter to produce a Mach--Zehnder fringe that is as symmetric as possible. The intensity ratio $I_2 /I_1$ was finally set to about 0.6, close to the theoretical value for zero relative ac Stark shift (0.56 for $^{87}$Rb and the detuning $\Delta$ = 2$\pi \times$ 1.07~GHz)\cite{malte_schmidt_mobile_2011}.
 
\section{Conclusion}
We have described a light-pulse atom interferometer that uses a continuous cold $^{87}Rb$ atomic source. Mach--Zehnder atomic interference fringes have been observed with this interferometer. Such interferometers may be used in a high precision gyroscope for its high bandwidth and technical flexibility. At present, we have achieved a small area of 0.07~$mm^2$ for the Sagnac loop and the corresponding scale factor of $ S=2k_{eff}L^2 /v_{z0}$ =194~rad/(rad/s), where L=9.5~mm and v$_{z0}$= 15~m/s. The deduced short-term sensitivity (1~s) of an atom gyroscope is about $7.8\times 10^{-5}$~$rad/s/\sqrt{{\rm Hz}}$, with a signal-to-noise ratio of 15.1 measured from the interference signal and a cutoff frequency of 190~Hz for the low-pass filter from sampling signals used in the experiment. With a theoretical bandwidth of $v_{z0}/2L$=790~Hz, the atom interferometer has a shot-noise-limited short-term sensitivity of $2\times 10^{-6}$~$rad/s/\sqrt{{\rm Hz}}$ for rotation sensing under current experimental conditions\cite{gustavson_rotation_2000}. The sensitivity can be improved further by establishing a larger area for the Sagnac loop, which is now limited to the size of the vacuum window.

The noise level of the atomic interference signal is currently limited by the intense background induced by the scattered light from the detection beam and Raman beams. The fluctuations of the interference signals mainly result from the jitter of the atomic flux and the Raman pulses. The stabilities of the cold atomic source and the Raman lasers (intensities and phase noise) need to be further optimized.

Compared with its pulsed counterparts, it is easier for a LPAI with a continuous source to be noise-compensated or loop-closed with the application of real-time frequency or phase modulation techniques\cite{gustavson_precision_2000,zhu_absolute_2008}. In the future, frequency and phase modulations are to be added to the spatially separated Raman pulses of our system.

\begin{acknowledgments}
This work was supported partly by the National Natural Science Foundation of China (Grant No. 61473166) and the
Major State Basic Research Development Program of China (973 Program) (Grant No. 2010CB922900). Yanying Feng thanks the China Scholarship Council  for support.
\end{acknowledgments}

\bibliography{references}

\end{document}